\documentclass[%
reprint,
superscriptaddress,
nofootinbib,
nobibnotes,
 amsmath,amssymb,
 aps,
 prl,
 longbibliography,
 twocolumn,
 floatfix
]{revtex4-2}
\usepackage[labelformat=simple]{subcaption}

\usepackage{epic}
\usepackage{float}
\usepackage[percent]{overpic}
\usepackage{graphicx, comment}
\usepackage{dcolumn}
\usepackage{bm}
\usepackage{xcolor}
\usepackage{hyperref}
\hypersetup{
    colorlinks=true,
    linkcolor=blue,
    filecolor=magenta,      
    urlcolor=cyan,
    pdftitle={numuCC0pNpPRL},
    pdfpagemode=FullScreen,
    }
\usepackage[normalem]{ulem}
\usepackage{amsfonts}
\usepackage{url}
\usepackage{graphicx}
\usepackage[newcommands]{ragged2e}

\usepackage{verbatim}

\begin{document}
\setlength{\parskip}{0pt}

\preprint{APS/123-QED}

\title{First simultaneous measurement of differential muon-neutrino charged-current cross sections on argon for final states with and without protons using MicroBooNE data}

\newcommand{\ANL}{Argonne National Laboratory (ANL), Lemont, IL, 60439, USA}
\newcommand{\Bern}{Universit{\"a}t Bern, Bern CH-3012, Switzerland}
\newcommand{\BNL}{Brookhaven National Laboratory (BNL), Upton, NY, 11973, USA}
\newcommand{\UCSB}{University of California, Santa Barbara, CA, 93106, USA}
\newcommand{\Cambridge}{University of Cambridge, Cambridge CB3 0HE, United Kingdom}
\newcommand{\CIEMAT}{Centro de Investigaciones Energ\'{e}ticas, Medioambientales y Tecnol\'{o}gicas (CIEMAT), Madrid E-28040, Spain}
\newcommand{\Chicago}{University of Chicago, Chicago, IL, 60637, USA}
\newcommand{\Cincinnati}{University of Cincinnati, Cincinnati, OH, 45221, USA}
\newcommand{\CSU}{Colorado State University, Fort Collins, CO, 80523, USA}
\newcommand{\Columbia}{Columbia University, New York, NY, 10027, USA}
\newcommand{\Edinburgh}{University of Edinburgh, Edinburgh EH9 3FD, United Kingdom}
\newcommand{\FNAL}{Fermi National Accelerator Laboratory (FNAL), Batavia, IL 60510, USA}
\newcommand{\Granada}{Universidad de Granada, Granada E-18071, Spain}
\newcommand{\Harvard}{Harvard University, Cambridge, MA 02138, USA}
\newcommand{\IIT}{Illinois Institute of Technology (IIT), Chicago, IL 60616, USA}
\newcommand{\Indiana}{Indiana University, Bloomington, IN 47405, USA}
\newcommand{\KSU}{Kansas State University (KSU), Manhattan, KS, 66506, USA}
\newcommand{\Lancaster}{Lancaster University, Lancaster LA1 4YW, United Kingdom}
\newcommand{\LANL}{Los Alamos National Laboratory (LANL), Los Alamos, NM, 87545, USA}
\newcommand{\Louisiana}{Louisiana State University, Baton Rouge, LA, 70803, USA}
\newcommand{\Manchester}{The University of Manchester, Manchester M13 9PL, United Kingdom}
\newcommand{\MIT}{Massachusetts Institute of Technology (MIT), Cambridge, MA, 02139, USA}
\newcommand{\Michigan}{University of Michigan, Ann Arbor, MI, 48109, USA}
\newcommand{\MSU}{Michigan State University, East Lansing, MI 48824, USA}
\newcommand{\Minnesota}{University of Minnesota, Minneapolis, MN, 55455, USA}
\newcommand{\Nankai}{Nankai University, Nankai District, Tianjin 300071, China}
\newcommand{\NMSU}{New Mexico State University (NMSU), Las Cruces, NM, 88003, USA}
\newcommand{\Oxford}{University of Oxford, Oxford OX1 3RH, United Kingdom}
\newcommand{\Pitt}{University of Pittsburgh, Pittsburgh, PA, 15260, USA}
\newcommand{\Rutgers}{Rutgers University, Piscataway, NJ, 08854, USA}
\newcommand{\SLAC}{SLAC National Accelerator Laboratory, Menlo Park, CA, 94025, USA}
\newcommand{\SDSMT}{South Dakota School of Mines and Technology (SDSMT), Rapid City, SD, 57701, USA}
\newcommand{\Maine}{University of Southern Maine, Portland, ME, 04104, USA}
\newcommand{\Syracuse}{Syracuse University, Syracuse, NY, 13244, USA}
\newcommand{\TelAviv}{Tel Aviv University, Tel Aviv, Israel, 69978}
\newcommand{\Tennessee}{University of Tennessee, Knoxville, TN, 37996, USA}
\newcommand{\UTA}{University of Texas, Arlington, TX, 76019, USA}
\newcommand{\Tufts}{Tufts University, Medford, MA, 02155, USA}
\newcommand{\UCL}{University College London, London WC1E 6BT, United Kingdom}
\newcommand{\VTech}{Center for Neutrino Physics, Virginia Tech, Blacksburg, VA, 24061, USA}
\newcommand{\Warwick}{University of Warwick, Coventry CV4 7AL, United Kingdom}
\newcommand{\Yale}{Wright Laboratory, Department of Physics, Yale University, New Haven, CT, 06520, USA}

\affiliation{\ANL}
\affiliation{\Bern}
\affiliation{\BNL}
\affiliation{\UCSB}
\affiliation{\Cambridge}
\affiliation{\CIEMAT}
\affiliation{\Chicago}
\affiliation{\Cincinnati}
\affiliation{\CSU}
\affiliation{\Columbia}
\affiliation{\Edinburgh}
\affiliation{\FNAL}
\affiliation{\Granada}
\affiliation{\Harvard}
\affiliation{\IIT}
\affiliation{\Indiana}
\affiliation{\KSU}
\affiliation{\Lancaster}
\affiliation{\LANL}
\affiliation{\Louisiana}
\affiliation{\Manchester}
\affiliation{\MIT}
\affiliation{\Michigan}
\affiliation{\MSU}
\affiliation{\Minnesota}
\affiliation{\Nankai}
\affiliation{\NMSU}
\affiliation{\Oxford}
\affiliation{\Pitt}
\affiliation{\Rutgers}
\affiliation{\SLAC}
\affiliation{\SDSMT}
\affiliation{\Maine}
\affiliation{\Syracuse}
\affiliation{\TelAviv}
\affiliation{\Tennessee}
\affiliation{\UTA}
\affiliation{\Tufts}
\affiliation{\UCL}
\affiliation{\VTech}
\affiliation{\Warwick}
\affiliation{\Yale}

\author{P.~Abratenko} \affiliation{\Tufts}
\author{O.~Alterkait} \affiliation{\Tufts}
\author{D.~Andrade~Aldana} \affiliation{\IIT}
\author{L.~Arellano} \affiliation{\Manchester}
\author{J.~Asaadi} \affiliation{\UTA}
\author{A.~Ashkenazi}\affiliation{\TelAviv}
\author{S.~Balasubramanian}\affiliation{\FNAL}
\author{B.~Baller} \affiliation{\FNAL}
\author{G.~Barr} \affiliation{\Oxford}
\author{D.~Barrow} \affiliation{\Oxford}
% ...

\collaboration{The MicroBooNE Collaboration}
\thanks{microboone\_info@fnal.gov}\noaffiliation


\date{February 29 2024}

\begin{abstract}
We report the first double-differential neutrino-argon cross section measurement made simultaneously for final states with and without protons for the inclusive muon neutrino charged-current interaction channel. The proton kinematics of this channel are further explored with a differential cross section measurement as a function of the leading proton's kinetic energy that extends across the detection threshold. These measurements use data collected with the MicroBooNE detector from 6.4$\times10^{20}$ protons on target from the Fermilab Booster Neutrino Beam with a mean neutrino energy of $\sim$0.8~GeV. Extensive data-driven model validation utilizing the conditional constraint formalism is employed. This motivates enlarging the uncertainties with an empirical reweighting approach to minimize the possibility of extracting biased cross section results. The extracted nominal flux-averaged cross sections are compared to widely used event generator predictions revealing severe mismodeling of final states without protons for muon neutrino charged-current interactions, possibly from insufficient treatment of final state interactions. These measurements provide a wealth of new information useful for improving event generators which will enhance the sensitivity of precision measurements in neutrino experiments.
\end{abstract}

\maketitle

Neutrino experiments that measure flavor oscillations as a function of neutrino energy aim to determine the neutrino mixing parameters and search for new physics beyond the Standard Model~\cite{PP_Review,Qian:2015waa,NOvA:2021nfi,T2K:2021xwb,DUNE2}. This requires precise mapping between reconstructed and true neutrino energy. The inclusive muon neutrino charged current ($\nu_\mu$CC) interaction channel, $\nu_\mu \text{N}\rightarrow\mu^-X$, where N is the struck nucleus and $X$ is the hadronic final state, is important for these measurements because it identifies the neutrino flavor with high purity and efficiency due to the sole requirement of detecting the muon. 

A number of these neutrino experiments utilize liquid argon time projection chambers (LArTPCs)~\cite{SBN,DUNE2}. These tracking calorimeters have low detection thresholds and excellent particle identification (PID) capabilities~\cite{LARTPC1,LARTPC2,LARTPC3,LARTPC4,LARTPC5}. LArTPCs enable the inclusive $\nu_\mu$CC channel to be divided into subchannels based on the composition of the final state, each having a different mapping between true and reconstructed neutrino energy. This improves the energy reconstruction and increases the sensitivity of precision measurements~\cite{WP_NuScat}.  

Maximizing the physics reach of LArTPCs requires neutrino-argon interaction modeling capable of describing all final state particles. Existing models are unable to describe data with such detail, necessitating large interaction modeling uncertainties~\cite{WP_NuScat,DUNE_sense}. This is unsurprising; theoretical models attempting to describe experimental observables must simultaneously account for multiple scattering mechanisms~\cite{evtoEev}, in-medium nuclear modifications to the fundamental neutrino interactions~\cite{ccinc,inc_semi_inc2}, and final-state interactions (FSI) involving the hadronic reaction products as they exit the nucleus~\cite{dytman_transparency}. The prominence of nuclear effects grows with the size of the target nucleus, further complicating the modeling of scattering for heavy nuclei like argon.

Efforts to simulate $\nu_\mu$CC interactions benefit from measurements that simultaneously probe the leptonic and hadronic kinematics. Building on previous MicroBooNE work~\cite{wc_1d_xs,wc_3d_xs,wc_elee}, and analogous to a similar measurement from T2K~\cite{t2k_cc0pi}, we report a double-differential measurement of the muon energy, $E_\mu$, and muon scattering angle with respect to the neutrino beam, $\cos\theta_\mu$, for the $\nu_\mu$CC channel split into final states with one or more protons (``Np" where N$\geq1$) and without protons (``0p"). An event is only included in the Np signal if the leading proton exiting the nucleus has kinetic energy above 35~MeV, which roughly corresponds to the proton tracking threshold in MicroBooNE~\cite{PRD}. 
The proton kinematics are further explored with a differential cross section measurement of the inclusive $\nu_\mu$CC channel (``Xp" where X$\geq0$) as a function of the leading proton's kinetic energy, $K_p$, that extends across the tracking threshold via the inclusion of a 0-35~MeV bin that includes events without a final state proton. A more expansive set of measurements employing the same analysis strategy can be found in~\cite{PRD}.

Additional motivation for these measurements comes from the fact that LArTPCs utilize the gap between the neutrino and shower vertices to differentiate electrons from photons. The absence of additional vertex activity, usually from protons, makes it difficult to determine if a gap is present. This impacts $\nu_e$CC selections, which are the signal in many oscillation measurements, through lower efficiencies and purities for $\nu_e$CC 0p events than $\nu_e$CC Np events~\cite{pelee,nue0pNp}. Since the $\nu_\mu$CC channel is essential in constraining the $\nu_e$CC prediction, improved modeling of 0p and Np final states for $\nu_\mu$CC is important. This need is highlighted by prior MicroBooNE results~\cite{wc_elee}, which observed an excess of $\nu_\mu$CC events at low reconstructed neutrino energies, potentially indicative of poor 0p cross section modeling. 

We utilize data collected with the MicroBooNE detector~\cite{uboone_detector} from an exposure of $6.4\times 10^{20}$ protons on target (POT) from the Booster Neutrino Beam (BNB) at a mean neutrino energy of $\sim$0.8~GeV~\cite{MiniBooNEFlux}. The detector is comprised of a TPC volume with an active mass of 85~tonnes of liquid argon, and an array of 32 photomultiplier tubes (PMTs). When an interaction occurs in the detector, scintillation light and ionization electrons are produced by the charged particles emanating from the interaction. The light is recorded by the PMTs, providing ns-scale timing used to reject background cosmic ray events that are out of time with the beam. The ionization electrons drift in a 273~V/cm electric field to three wire readout planes which record charge distributions used for calorimetry and three-dimensional (3D) mm-scale imaging. 

Event reconstruction, calorimetry, and PID are performed with the Wire-Cell topographical 3D image processing algorithm~\cite{wc_reco}. Wire charge distributions are first deconvolved from the detector response by a TPC signal processing algorithm~\cite{tpc_signal_proc,sig_proc_1,sig_proc_2}.  Wire-Cell uses the deconvolved readouts to reconstruct 3D images without topological assumptions about the source of activity as ``tracks", which leave continuous energy depositions, or electromagnetic ``showers"~\cite{WC3D}, which deposit more charge perpendicular to their trajectory. A many-to-many TPC-charge to PMT-light matching algorithm is used for further cosmic ray rejection~\cite{wire-cell-uboone}. 

Particle identification starts with finding kinks in the selected group of charge activity to identify tracks~\cite{WC3D}. Candidate neutrino vertices and final state particles are identified concurrently based on $dQ/dx$, topology, and allowed particle relationships. A final neutrino vertex is chosen by a $\texttt{SparseConvNet}$ deep neural network~\cite{SparseConvNet}. Proton and muon candidates are distinguished based on characteristic differences in their $dQ/dx$ profile~\cite{PRD} using a test statistic constructed from a Kolmogorov–Smirnov shape comparison score and the normalization to the median $dQ/dx$ of protons.

Two methods are used to calculate the energy of track-like particles; range and summation of $dE/dx$. Between the two methods, the energy resolution is $\sim$10\% for muons at all energies and $\sim$8\% for protons below 200~MeV, above this the resolution degrades to $\sim$25\% due to the increased probability of re-scattering~\cite{PRD}. The range method is used for stopping tracks and is based on the NIST PSTAR database~\cite{pstar}. This method estimates the energy of tracks with minimal bias~\cite{PRD}. Summation of $dE/dx$ is used for tracks that are shorter than 4~cm, exit the active volume, have a ``wiggled” topology indicative of many small angle deflections~\cite{WC3D}, or emit $\delta$ rays. To calculate the kinetic energy, this method converts $dQ/dx$ to $dE/dx$ with an effective recombination model then sums $dE/dx$ for each $\sim$6~mm segment of the track. The $dE/dx$ method underestimates the energy by $\sim$10\%~\cite{WC3D,uboone_energy_cal}, but this is incorporated into the detector model and appears consistent in data and simulation~\cite{wc_elee}.   

The $\nu_\mu$CC event selection is identical to that of previous MicroBooNE work~\cite{wc_elee,wc_1d_xs,wc_3d_xs,PRD}. It utilizes the ``generic neutrino selection"~\cite{cosmic_reject} as a preselection, which reduces cosmic ray contamination down to 15\%. Backgrounds are further rejected with a boosted decision tree (BDT) trained using the $\texttt{XGBoost}$ library~\cite{xgboost} on a set of background taggers, which are variables designed to characterise non-$\nu_\mu$CC events. The efficiency (purity) of the fully inclusive $\nu_\mu$CC selection is 68\% (92\%), with backgrounds predominantly coming from neutral current $\pi^\pm$ events. The selection is further divided into 0p and Np selections based on the reconstructed leading proton's kinetic energy, $K_p^{rec}$. The Np selection contains events in which there is at least one proton with $K_p^{rec}>35$~MeV. The 0p selection contains all other events. True Np events are analogously defined as having a proton with true $K_p>35$~MeV and true 0p events are defined as having either zero final-state protons, or no proton with $K_p>35$~MeV. The Np selection has a 49\% efficiency for true Np events and a high purity of 95\% due to low contamination from non-$\nu_\mu$CC and 0p events. The 0p selection has an efficiency of 53\% for true 0p events. The larger number of Np events, which outnumber 0p events $\sim$7:1, increases the prominence of Np events in the 0p selection, reducing the fraction of true 0p events in the 0p selection to 32\%~\cite{PRD}.

The 0p and Np cross sections are extracted simultaneously. This allows the number of true Np events in the 0p selection to be predicted based on the observation of the Np selection (and vice versa). The same strategy is employed in other MicroBooNE work~\cite{PRD,nue0pNp}. Simultaneous extraction requires unfolding
\begin{equation} \label{eq:master}
 \begin{pmatrix} M_{0p} \\  M_{Np} \end{pmatrix} = \begin{pmatrix} R_{0p0p} &  R_{0pNp} \\ R_{Np0p} & R_{NpNp} \end{pmatrix} \cdot \begin{pmatrix} S_{0p} \\  S_{Np} \end{pmatrix} + \begin{pmatrix} B_{0p} \\  B_{Np} \end{pmatrix}, 
\end{equation}
where $M$ is the reconstructed distribution, $S$ is the differential cross section to be extracted, $B$ is the distribution of background events that are not part of the inclusive $\nu_\mu$CC channel, and $R$ is the response matrix describing the mapping between the true and reconstructed distributions. The first (second) index on $R$ corresponds to the reconstructed (true) proton multiplicity.

The Wiener-SVD unfolding technique~\cite{WSVD} is used to extract nominal flux-averaged cross section results~\cite{flux_uncertainty_rec}. This method returns a regularized unfolded cross section and corresponding covariance matrix, $V_S$, which describes the uncertainties and bin-to-bin correlations on the result. The form of Eq.~(\ref{eq:master}) allows the unfolding to account for correlations between the 0p and Np channels~\cite{PRD}. An additional smearing matrix, $A_C$, that captures the bias induced by regularization is also obtained in the unfolding. Cross section predictions should be multiplied by $A_C$ when compared to the unfolded result. The extracted cross sections, $A_C$, and $V_S$ which is obtained via blockwise unfolding to preserve inter-variable correlations~\cite{GardinerXSecExtract,PRD}, can be found in the Supplemental Material.

Monte Carlo (MC) simulations are used to estimate $R$ and part of $B$. The neutrino flux is modeled with the $\texttt{Geant4}$ simulation of the BNB from MiniBooNE~\cite{Geant4,MiniBooNEFlux}. Neutrino-argon interactions are simulated with the $\texttt{G18\_10a\_02\_11a}$ configuration of the $\texttt{GENIE v3.0.6}$ event generator ~\cite{GENIE} tuned to $\nu_\mu$CC data without final state pions from T2K~\cite{T2KTuneData} by reweighting based on two CC quasi-elastic and CC meson-exchange-current parameters~\cite{uboonetune}. The resulting prediction is referred to as the ``MicroBooNE Tune". Final state particles are propagated through a detector simulation using the $\texttt{Geant4}$ toolkit $\texttt{v4\_10\_3\_03c}$~\cite{Geant4} and processed using the $\texttt{LArSoft}$~\cite{larsoft} framework. The simulated TPC and PMT waveforms are overlaid on beam-off data to provide an accurate description of cosmic ray activity. These overlaid MC samples are processed like real data and used to estimate $R$ and $B$.

Uncertainties are estimated with covariance matrices calculated from a multi-universe approach as in~\cite{PRD}. The total covariance matrix, $V^{\mathrm{sys}} =  V_{\mathrm{flux}} + V_{\mathrm{reint}} + V_{\mathrm{xs}} + V_{\mathrm{det}} + V_{\mathrm{MC}}^{\mathrm{stat}} + V_{\mathrm{dirt}} + V_{\mathrm{POT}} + V_{\mathrm{Target}} + V_{\mathrm{rw}}$, is given by the sum of the covariance matrices calculated for each systematic uncertainty described below.

Uncertainty in the neutrino flux~\cite{MiniBooNEFlux} is contained in $V_{\mathrm{flux}}$. The flux contributes 5-10\% uncertainty to the cross section results and is often the dominant systematic for the Np cross section measurements. Neutrino-argon cross section uncertainties are accounted for in $V_{\mathrm{xs}}$~\cite{uboonetune} and contribute $\sim$5$\%$ uncertainty. In $V_{\mathrm{reint}}$, uncertainties on reinteractions of final state particles outside the nucleus are accounted for. These are estimated using the $\texttt{Geant4Reweight}$~\cite{geantrw} package and are relatively small except at high $K_p$ where reinteractions occur for $>75\%$ of protons. The flux, cross section, and re-interaction uncertainties are estimated with the multi-sim technique~\cite{multisim}.

Detector response uncertainties~\cite{detvar} are contained in $V_{\mathrm{det}}$. These are the biggest sources of uncertainty for 0p events, typically ranging from 5-15\% compared to $\sim$5\% for Np, and are larger at high energies and backward scattering angles. To evaluate these uncertainties, a detector model parameter is varied by $1\sigma$ and bootstrapping is used to estimate the impact of this variation and form $V_{\mathrm{det}}$~\cite{wc_elee,PRD}. A Gaussian process regression smoothing algorithm~\cite{gpr1,gpr2,gpr3} is implemented to prevent an overestimation of detector systematics due to statistical fluctuations~\cite{wc_3d_xs,PRD}.

Flat 50\%, 2\% and 1\% uncertainties on neutrino interactions outside the detector, POT counting, and the number of target nuclei are contained in $V_{\mathrm{dirt}}$, $V_{\mathrm{POT}}$ and $V_{\mathrm{Target}}$, respectively. Their impact on the extracted cross sections is small.

A data-driven model validation procedure is employed to detect mismodeling that may bias the extracted cross section results. As described in~\cite{PRD}, this relies on the conditional constraint formalism~\cite{cond_cov} to increase the stringency of the validation. These constraints leverage correlations between variables and channels arising from shared physics modeling to update the model prediction and reduce its uncertainties based on data observations. The unfolding does not utilize these constraints; they are only for model validation. To be validated, the model is required to describe the data at the $2\sigma$ level. This is evaluated with $\chi^2$ goodness of fit (GoF) tests interpreted by using the number of degrees of freedom, $ndf$, which corresponds to the number of bins, to obtain $p$-values. 

A kinematic variable that is relatively well understood and reconstructed can be validated by directly comparing the model prediction to the data in the phase space relevant to the unfolding. The muon kinematics for events fully contained (FC) within the detector fit this criteria, and are validated with a GoF test on the $E_\mu^{rec}$ distributions in $\cos\theta_\mu^{rec}$ slices used for the cross section extraction. These tests yield $p$-values of 0.45 and 0.98 for the 0p and Np distributions, respectively. The model passes validation in these tests. For the partially contained (PC) muon kinematics, the modeling of activity outside the active detector volume, which cannot be reconstructed, must also be validated. This is done by using the 0p and Np FC muon kinematics to constrain the PC distributions. These tests, found in the Supplemental Material of~\cite{PRD}, result in $p$-values of 0.84 for 0p and 0.99 for Np. This indicates that the overall model adequately describes the 0p and Np $E_\mu^{rec}$ distributions in $\cos\theta_\mu^{rec}$ slices for PC events.  

\begin{figure}[h!]
  \begin{subfigure}[t]{\linewidth}
  \includegraphics[width=\linewidth]{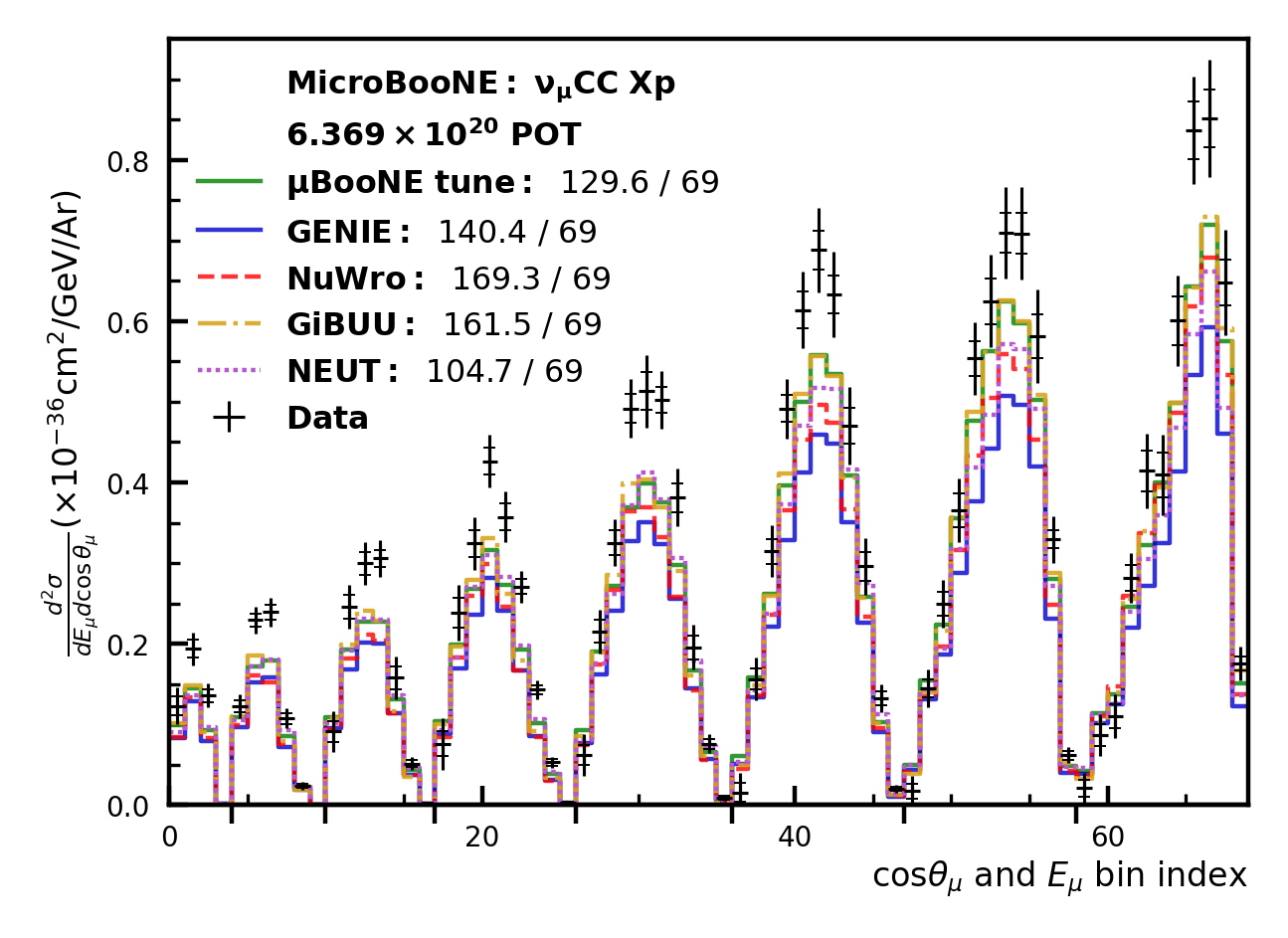}
  \vspace{-63mm}\caption{\hspace*{55mm}\label{CosThetaMuEmu_xs_Xp}}
  \end{subfigure}\vspace*{-3.2mm}
 \begin{subfigure}[t]{\linewidth}
 \includegraphics[width=\linewidth]{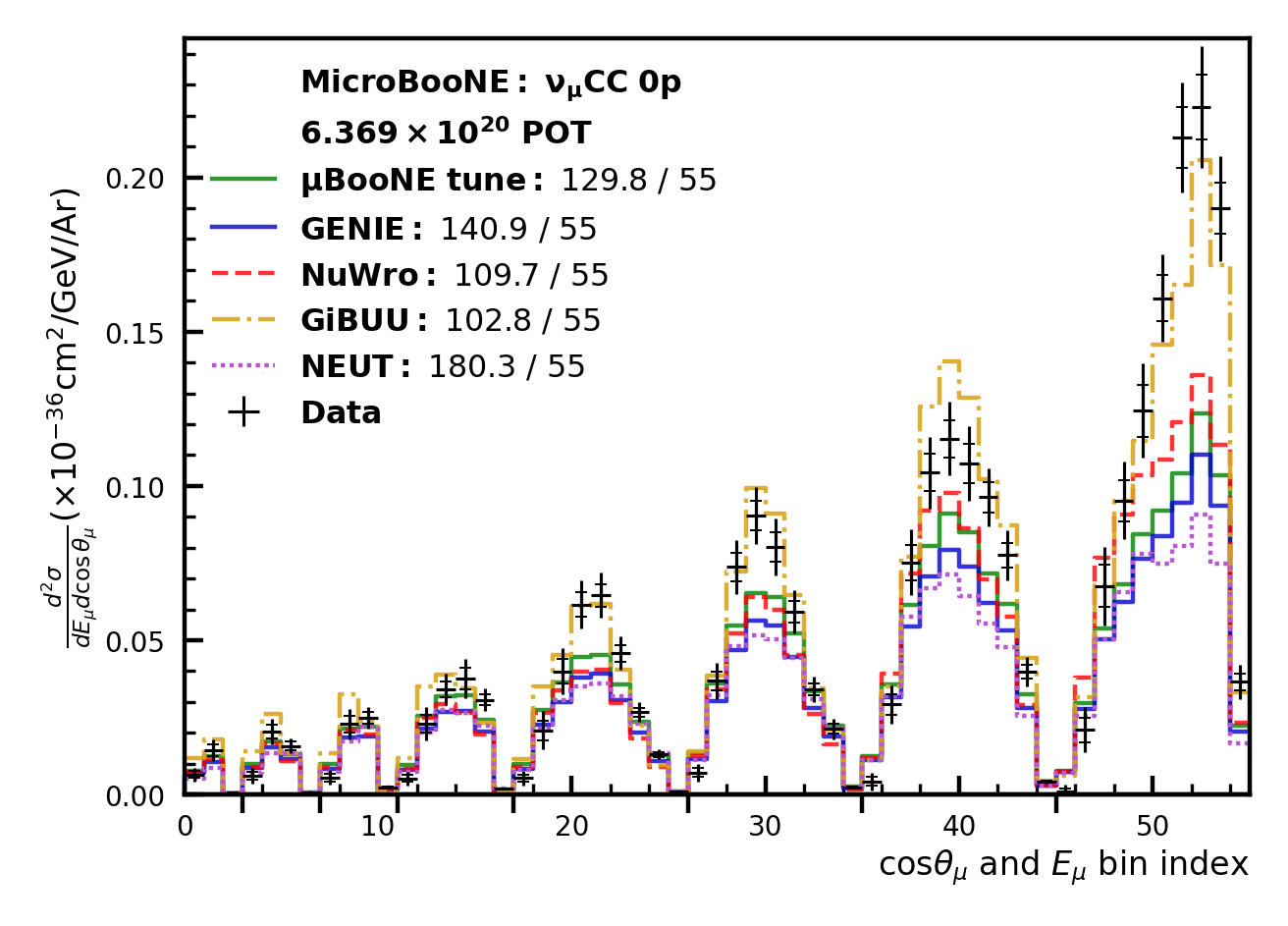}
  \vspace{-62mm}\caption{\hspace*{54mm}\label{CosThetaMuEmu_xs_0p}}
  \end{subfigure}\vspace*{-3.2mm}
 \begin{subfigure}[t]{\linewidth}
 \includegraphics[width=\linewidth]{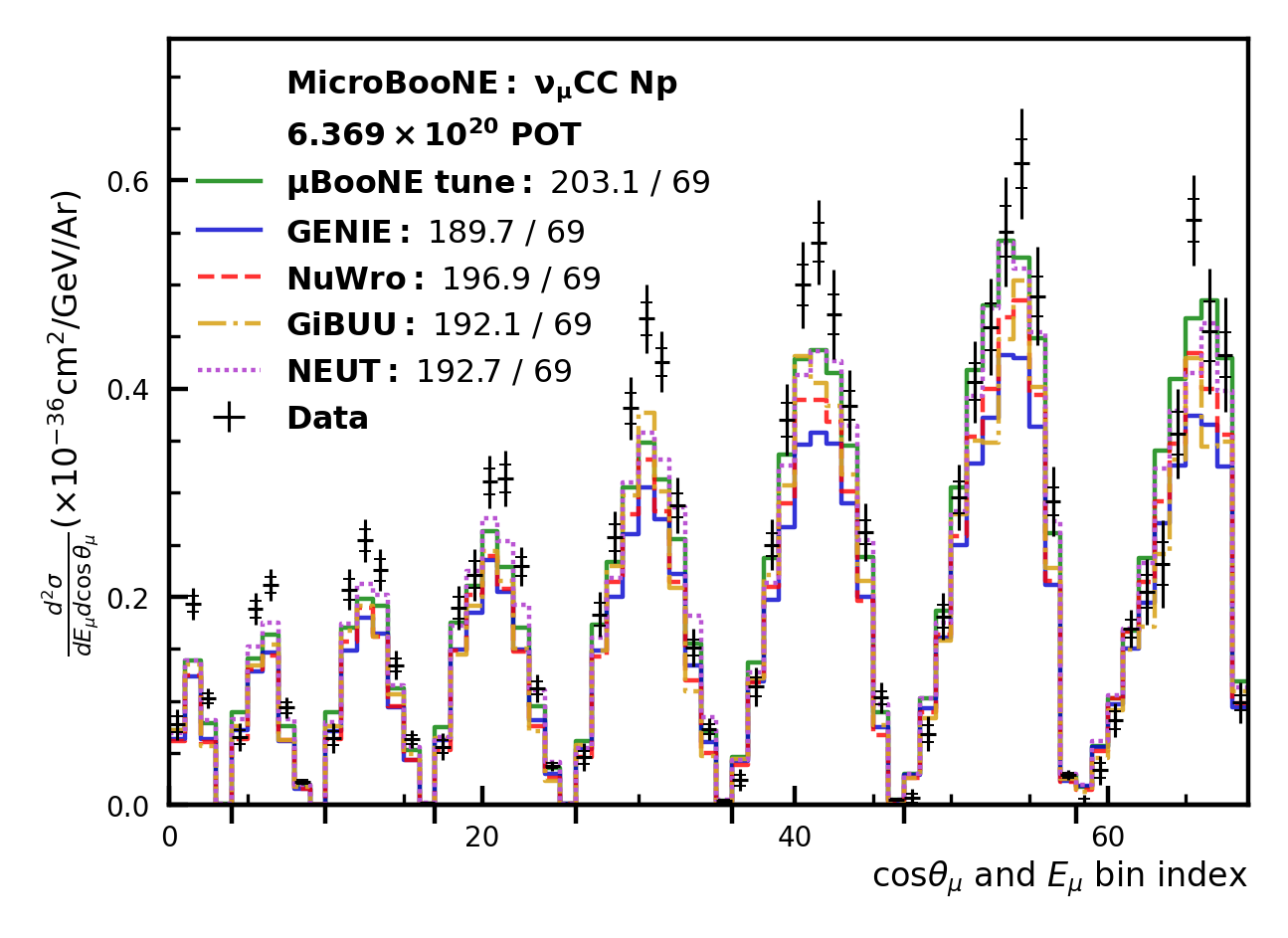}
  \vspace{-63mm}\caption{\hspace*{55mm}\label{CosThetaMuEmu_xs_Np}} 
  \end{subfigure}\vspace*{-1.2mm}
  \caption{\justifying The unfolded double-differential $\cos\theta_\mu$ and $E_\mu$ cross section for the $\nu_\mu$CC channel. The bins are in angular slices indicated by the downwards ticks on the x-axis; their edges are $\{-1,-0.5,0,0.3,0.5,0.7,0.8,0.9,1\}$.~The bins increase in energy along each slice. Binning details are in the Supplemental Material. The inclusive Xp result is shown in (a) and the simultaneously extracted 0p and Np results are shown in (b) and (c), respectively. The inner (outer) error bars represent the data statistical (total) uncertainty. Generator predictions correspond to the colored lines with $\chi^2/ndf$ displayed in the legend. These predictions are smeared with $A_C$ obtained in the unfolding.}
\label{CosThetaMuEmu_xs}
\end{figure}

When the $K_p^{rec}$ distribution is constrained by the 0p and Np muon kinematics the resulting $p$-value is $3\times10^{-5}$, indicating that the model is unable to describe the observed proton kinematics. The discrepancy appears at low $K_p^{rec}$ where modeling becomes challenging due to the prominence of FSI and nuclear effects. Appropriate uncertainties on $K_p^{rec}$ are important for the $K_p$ differential cross section measurement and the split into 0p and Np subchannels. Thus, this shortcoming of the model motivates expanding the uncertainties to mitigate the possibility of extracting biased results. The additional uncertainty was obtained empirically by estimating a true $K_p$ distribution for FC signal events passing the generic neutrino selection by unfolding the FC $K_p^{rec}$ data using the constrained prediction and only statistical uncertainties. The ratio of the unfolded distribution to the constrained signal prediction defines a reweighting function. This is applied to all $\nu_\mu$CC signal events and is treated as a $1\sigma$ deviation from the original prediction due to a cross section effect allowing $V_{\mathrm{rw}}$ to be calculated identically to $V_{\mathrm{xs}}$~\cite{PRD}. Including $V_{\mathrm{rw}}$ reduces the $p$-value for the $K_p^{rec}$ distribution to 0.82 after constraint, enabling the extraction of the desired cross sections.

The extracted nominal flux-averaged cross sections are compared to event generator predictions from $\texttt{GENIE v3.0.6 G18\_10a\_02\_11a}$ ($\texttt{GENIE}$)~\cite{GENIE}, the ``MicroBooNE tune" $\texttt{GENIE}$ configuration ($\mu\texttt{BooNE}$ tune)~\cite{uboonetune}, $\texttt{NuWro 21.02}$ ($\texttt{NuWro}$)~\cite{nuwro}, $\texttt{GiBUU 2023}$ ($\texttt{GiBUU}$)~\cite{gibuu2},  and $\texttt{NEUT 5.4.0.1}$ ($\texttt{NEUT}$)~\cite{neut}. These were processed with $\texttt{NUISANCE}$~\cite{NUISANCE}, do not include theoretical uncertainties, and are smeared with $A_C$ obtained from unfolding. Agreement between the data and each prediction is quantified by $\chi^2/ndf$ values calculated with uncertainties according to $V_S$ and $ndf$ corresponding to the number of bins.

The double-differential $\cos\theta_\mu$ and $E_\mu$ cross section results are shown in Fig.~\ref{CosThetaMuEmu_xs} as a function of bin index, which are in angular slices ranging from backwards on the left to forward on the right and increasing in energy along each slice. Binning details are in the Supplemental Material. In all three channels, the generators tend to underpredict the peak of the $E_\mu$ distribution, with the only exception being for $\texttt{GiBUU}$ in the 0p channel, where its prediction shows good normalization agreement around the peak. 

Figure~\ref{CosThetaMuEmu_xs_Xp} shows the fully inclusive Xp result. The $\chi^2$ values indicate that $\texttt{NEUT}$ best describes this data. The 0p result is shown in Fig.~\ref{CosThetaMuEmu_xs_0p}. At forward angles, $\texttt{GiBUU}$ describes this result better than the other generators, which significantly underpredict the cross section. Agreement at backwards angles is more comparable, but $\texttt{GiBUU}$ still has the lowest $\chi^2/ndf$ over all bins. $\texttt{NEUT}$, though offering the best description of the inclusive channel, shows the largest discrepancy with the 0p data. Figure~\ref{CosThetaMuEmu_xs_Np} shows the Np result. The $\chi^2$ values for Np are comparable between the different generators. Simultaneously extracting the 0p and Np cross sections allows the results to be examined concurrently with a $\chi^2$ calculated across all bins. These $\chi^2/ndf$ values are: 287.5/124 for the $\mu\texttt{BooNE}$ tune, 266.3/124 for $\texttt{GENIE}$, 263.7/124 for $\texttt{NuWro}$, 298.8/124 for $\texttt{NEUT}$, and 249.8/124 for $\texttt{GiBUU}$. These indicate that $\texttt{GiBUU}$ best describes the data when the proton content of the hadronic final state is examined in more detail.

\begin{figure}
\centering
  \includegraphics[width=\linewidth]{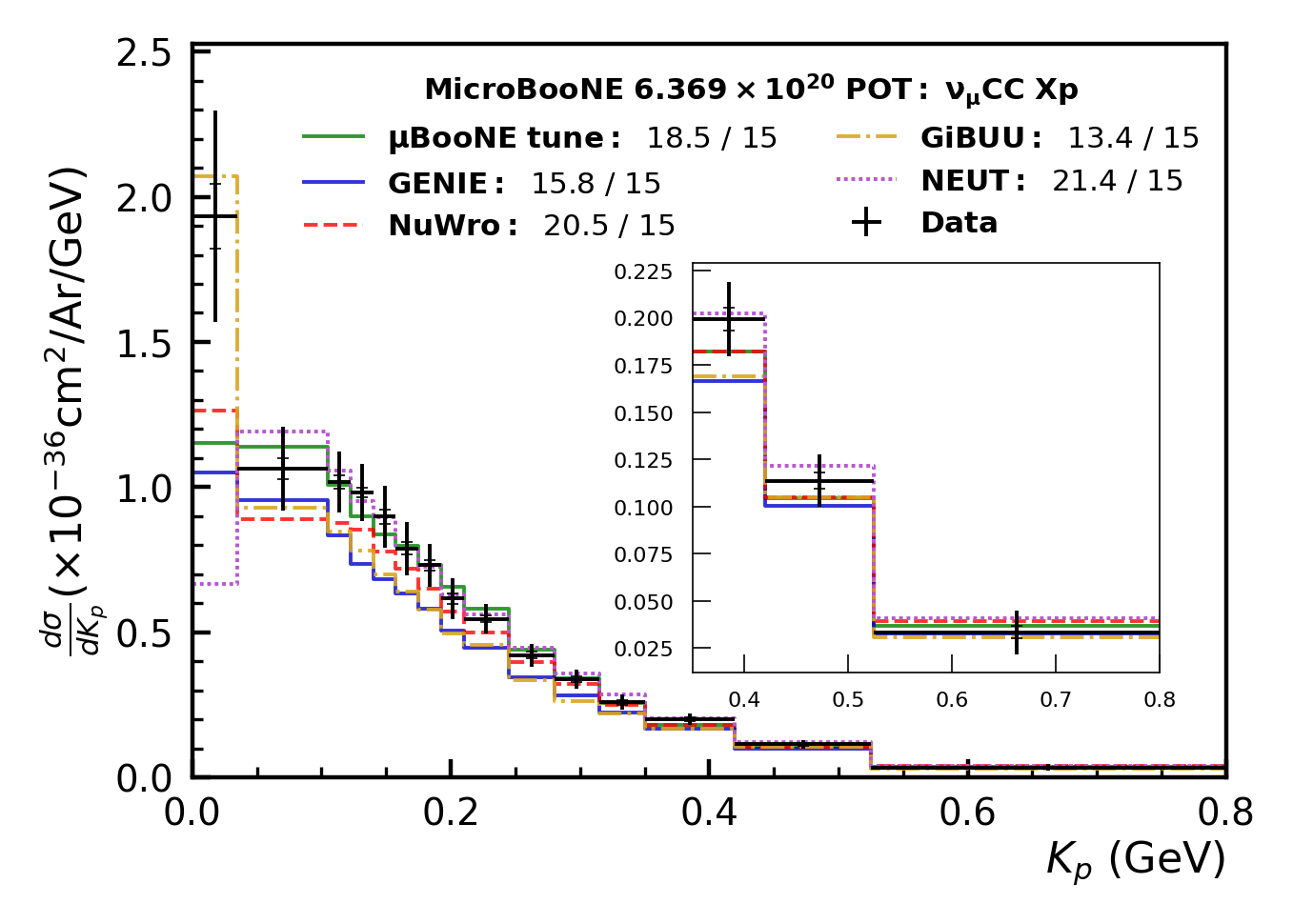}
    \put(-200.8,25.5){\textcolor{black}{\linethickness{0.8pt}\dashline[70]{7}(0,0)(0,142)}}
  \caption{\justifying The unfolded differential cross section for the inclusive $\nu_\mu$CC channel as a function of the leading proton's kinetic energy. The dashed line indicates the 35~MeV tracking threshold, below which is a single bin that includes events without protons. The inset magnifies the last three bins. See Fig.~\ref{CosThetaMuEmu_xs} for style details.}
\label{Kp_xs}
\end{figure}

The differential cross section for the inclusive channel as a function of $K_p$ is shown in Fig.~\ref{Kp_xs}. The first bin extends from 0-35~MeV and includes events without a final state proton; it is equivalent to the 0p signal definition. The generator predictions diverge at low energies, particularly for the 0p bin where only $\texttt{GiBUU}$ is able to describe the data. This gives $\texttt{GiBUU}$ the lowest $\chi^2$ despite its underprediction of the data at moderate-to-high energies. Similar underprediction is seen for $\texttt{GENIE}$ and $\texttt{NuWro}$ in this region. $\texttt{NEUT}$ and the $\mu\texttt{BooNE}$ tune describe moderate-to-high energies well, but $\texttt{NEUT}$ also significantly underpredicts the 0p bin, increasing its $\chi^2$.

Comparing the 0p and Np results to the Xp result in Fig.~\ref{CosThetaMuEmu_xs} demonstrates how a model that does well for inclusive scattering may not also be able to describe the hadronic final state. $\texttt{NEUT}$ stands out in its relatively good description of Xp but not of 0p due to a significant underprediction of the 0p cross section. This is possibly attributable to the way $\texttt{NEUT}$ treats binding energy for nucleon FSI. $\texttt{NEUT}$ assigns nucleons an effective mass when propagating them through the nucleus and only allows interactions if the total energy is twice the energy of the free nucleon mass~\cite{neut}. This reduces the strength of FSI for low energy nucleons~\cite{dytman_transparency} leading to a sharp drop-off in the cross section at low $K_p$ and the low 0p cross section prediction that agrees poorly with the data.

Unlike $\texttt{NEUT}$, $\texttt{GiBUU}$ describes the data relatively well when the $\nu_\mu$CC channel is split into final states with and without protons. This can possibly be attributed to its implementation of FSI with a transport model, which includes a binding potential that treats ``target" and ``ejected" nucleons identically. The ``ejected" nucleons are propagated on realistic trajectories through the residual nucleus according to a potential consistent with the initial interaction~\cite{gibuu,gibuu2}. 
This is absent in other generators, which propagate ``ejected" nucleons on straight lines and decouple the initial interaction from the FSI. After the initial interaction, an outgoing nucleon may repeatedly collide with other nucleons, depleting the initial nucleon of its energy and shifting the $K_p$ distribution towards smaller values~\cite{dytman_transparency,gibuu_minerva,nuwro_fsi2,gibuu3,gibuu}. It is plausible that the more self-consistent treatment of FSI in $\texttt{GiBUU}$ better captures this effect. This hypothesis is consistent with the way $\texttt{GiBUU}$ better describes the 0p final states in Fig.~\ref{CosThetaMuEmu_xs}, and has the only prediction that mirrors the sharp peak in the data at the lowest $K_p$ in Fig.~\ref{Kp_xs}. Additional 0p and Np results in~\cite{PRD} and measurements of transverse kinematic imbalance variables~\cite{TKI} sensitive to FSI modeling in~\cite{afro,afroPRL} support this hypothesis. Of course, different modeling of the initial neutrino-nucleon interaction could also play a role, especially given that none of the generators adequately describe the inclusive measurement shown in Fig.~\ref{CosThetaMuEmu_xs_Xp}. 

In summary, we report differential cross-section measurements of the $\nu_\mu$CC channel that probe the phase space of lepton and hadronic kinematics. An underprediction of the cross section for final states without protons is observed for all event generator predictions except $\texttt{GiBUU}$, which offers a significantly better description of the data possibly due to its more sophisticated treatment of final state interactions. These measurements provide new information to stimulate further improvement of models and generators to match the precision required for future neutrino oscillation measurements and beyond the Standard Model searches.

\begin{acknowledgments}
This document was prepared by the MicroBooNE collaboration using the resources of the Fermi National Accelerator Laboratory (Fermilab), a U.S. Department of Energy, Office of Science, HEP User Facility. Fermilab is managed by Fermi Research Alliance, LLC (FRA), acting under Contract No. DE-AC02-07CH11359.  MicroBooNE is supported by the following: the U.S. Department of Energy, Office of Science, Offices of High Energy Physics and Nuclear Physics; the U.S. National Science Foundation; the Swiss National Science Foundation; the Science and Technology Facilities Council (STFC), part of the United Kingdom Research and Innovation; the Royal Society (United Kingdom); the UK Research and Innovation (UKRI) Future Leaders Fellowship; and the NSF AI Institute for Artificial Intelligence and Fundamental Interactions. Additional support for the laser calibration system and cosmic ray tagger was provided by the Albert Einstein Center for Fundamental Physics, Bern, Switzerland. We also acknowledge the contributions of technical and scientific staff to the design, construction, and operation of the MicroBooNE detector as well as the contributions of past collaborators to the development of MicroBooNE analyses, without whom this work would not have been possible. For the purpose of open access, the authors have applied a Creative Commons Attribution (CC BY) public copyright license to any Author Accepted Manuscript version arising from this submission.
\end{acknowledgments}

\bibliography{prl.bib}

\end{document}